\documentclass[trackchanges,twocolumn,times]{aastex63}

\shorttitle{Solar wind IFIP}                 

\shortauthors{Brooks et al. }
\journalinfo{To be published in the Astrophysical Journal Letters}

\usepackage{grffile}

\begin{document}

\title{Detection of stellar-like abundance anomalies in the slow solar wind}

\author[0000-0002-2189-9313]{David H.\ Brooks}
\affil{College of Science, George Mason University, 4400 University Drive, Fairfax, VA 22030, USA}

\author[0000-0000-0000-0000]{Deborah Baker}
\affil{University College London, Mullard Space Science Laboratory, Holmbury St. Mary, Dorking, Surrey, RH5 6NT, UK}

\author[0000-0002-2943-5978]{Lidia van Driel-Gesztelyi}
\affil{University College London, Mullard Space Science Laboratory, Holmbury St. Mary, Dorking, Surrey, RH5 6NT, UK}
\affil{LESIA, Observatoire de Paris, Universit\'e PSL, CNRS, Sorbonne Universit\'e, Univ. Paris Diderot, Sorbonne Paris Cit\'e, 5 place Jules Janssen, 92195 Meudon, France}
\affil{Konkoly Observatory, Research Centre for Astronomy and Earth Sciences, Konkoly Thege \'ut 15-17., H-1121, Budapest, Hungary}

\author[0000-0001-6102-6851]{Harry P. Warren}
\affil{Space Science Division, Naval Research Laboratory, Washington, DC 20375, USA}

\author[0000-0003-2802-4381]{Stephanie L. Yardley}
\affil{University College London, Mullard Space Science Laboratory, Holmbury St. Mary, Dorking, Surrey, RH5 6NT, UK}

\begin{abstract}
The elemental composition of the Sun’s hot atmosphere, the corona, shows a distinctive pattern that is different than the underlying surface, or photosphere \citep{Pottasch1963}. Elements that are easy to ionize in the chromosphere are enhanced in abundance in the corona compared to their photospheric values. A similar pattern of behavior is often observed in the slow speed ($<$ 500\,km s$^{-1}$) solar wind \citep{Meyer1985}, and in solar-like stellar coronae \citep{Drake1997}, while a reversed effect is seen in M-dwarfs \citep{Liefke2008}. Studies of the inverse effect have been hampered in the past because only unresolved (point source) spectroscopic data were available for these stellar targets. Here we report the discovery of several inverse events observed in-situ in the slow solar wind using particle counting techniques. These very rare events all occur during periods of high solar activity that mimic conditions more widespread on M-dwarfs. The detections allow a new way of connecting the slow wind to its solar source, and are broadly consistent with theoretical models of abundance variations due to chromospheric fast mode waves with amplitudes of 8--10\,km s$^{-1}$; sufficient to accelerate the solar wind. The results imply that M-dwarf winds are dominated by plasma depleted in easily ionized elements, and lend credence to previous spectroscopic measurements.
\end{abstract}

\section{Introduction}
The enhancement of low first ionization potential (FIP; less than $\sim$10 eV) elements by factors of 2--4 in the solar corona has become known as the FIP effect \citep{Feldman1992}. A combination of remote sensing (spectroscopic) measurements of abundances in the solar atmosphere, and in-situ (particle counting) measurements in the solar wind, has led to a vast literature describing elemental composition in different solar structures \citep{Feldman2000}, their temporal behavior in different features and dynamic events \citep{Sheeley1995,Widing1997,Warren2016}, their use as tracers of a connection to the solar wind \citep{Brooks2011,Brooks2015}, or solar energetic particles \citep{Brooks2021}, and the development of theoretical models of abundance anomalies \citep{Laming2004}. These studies are complemented by measurements in the range of conditions observed in the coronae of low activity solar-like stars.

An inverse FIP (IFIP) effect was detected as a depletion of Fe in the coronae of active stars compared to their photospheres in observations made by the Advanced Satellite for Cosmology and Astrophysics (ASCA) and the Extreme UltraViolet Explorer (EUVE); see \cite{Antunes1994,Stern1995,Rucinski1995,Schmitt1996}, and further references in the review by \cite{Drake2002}. For discussion of more recent observations of the IFIP effect see also \cite{Wood2010} and \cite{Testa2015}. Furthermore, the IFIP effect was recently shown to be present also on stars with higher surface temperatures, if they have evolved off the main sequence and have some indicators of magnetic activity e.g. a high rotation rate, or X-ray luminosity \citep{Seli2022}. But studies of the IFIP effect are relatively less developed, because until recently it had only been observed on unresolved stellar targets. In 2015, however, the IFIP effect was discovered on the Sun \citep{Doschek2015}, associated with sunspots during solar flares, allowing the first detailed studies of the process at high spatial and temporal resolution. One possibility is that the IFIP effect observed in M-dwarfs is due to enhanced levels of flaring activity. The effect is present in their coronae all the time, however, even when they are not flaring, or at least not producing large flares. Furthermore, observations of elemental abundance changes during flares on several stars which have IFIP composition or metal depleted coronae, show that the plasma composition evolves towards photospheric during the event \citep{Stern1992,Mewe1997,Gudel1999,Liefke2010}. This is consistent with the new solar observations, which suggest that IFIP composition plasma is present in cases where, in the presence of emerging flux, magnetically complex sunspots collide, and the flare only acts to reveal (not cause) it by producing high 
\begin{figure}[h]
\centering
\includegraphics[width=0.5\textwidth]{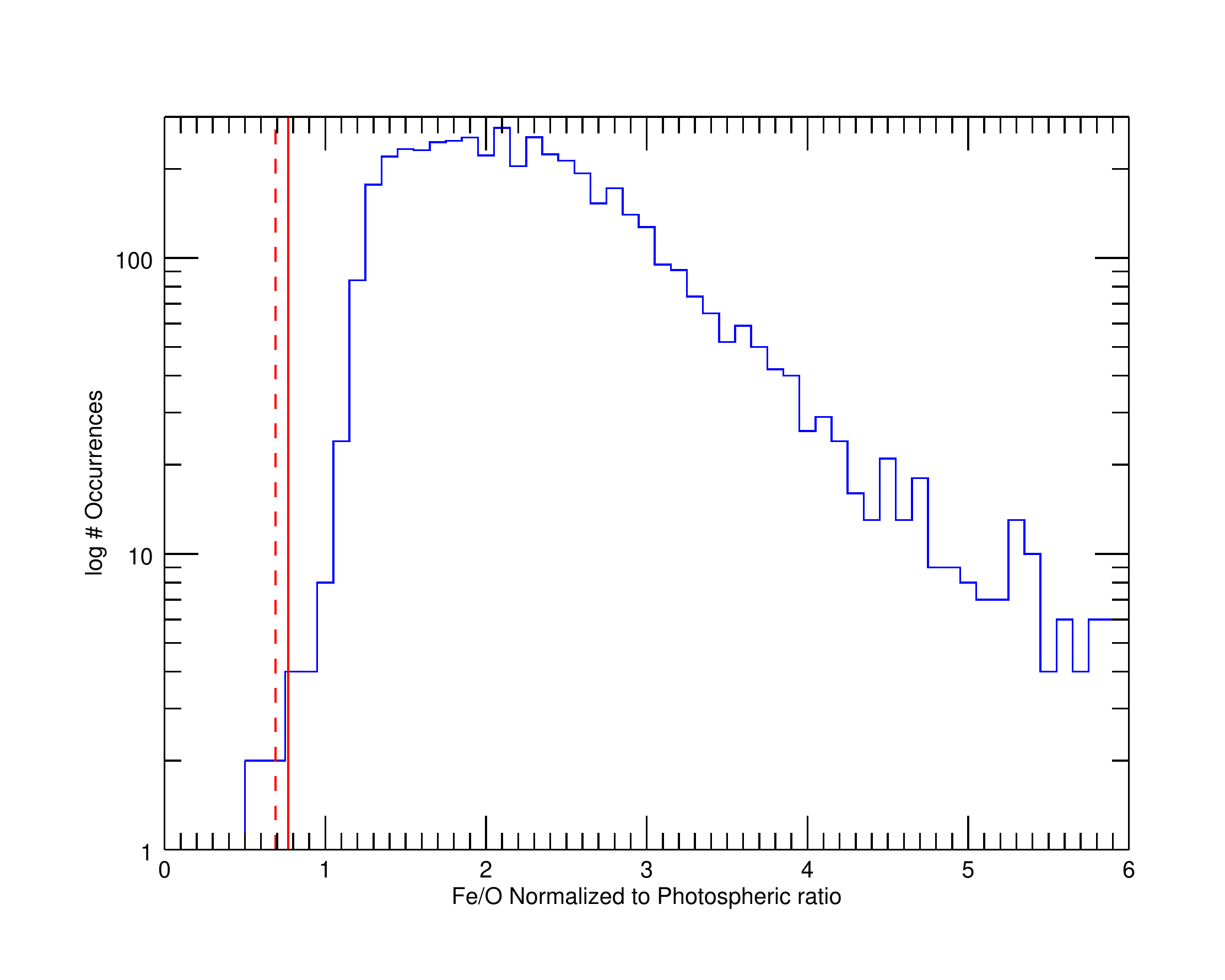}
\caption{ ACE/SWICS measurements of Fe/O normalized to their photospheric ratio. The data were obtained between 1998 February and 2011 August. The red solid line indicates the level we have defined for IFIP detection. Whether the measurements fall below this level depends on the assumed photospheric Fe/O abundance and these vary in the literature, so for reference we also show the threshold that would need to be reached if a more extreme lower value is chosen (see section \ref{Methods}).
}
\label{fig1}
\end{figure}
temperatures \citep{Baker2020}. Highly magnetically complex regions are more likely to exist, and dominate the emission, on the more active solar-like stars. 

\begin{figure*}
\centering
\includegraphics[width=1.0\textwidth]{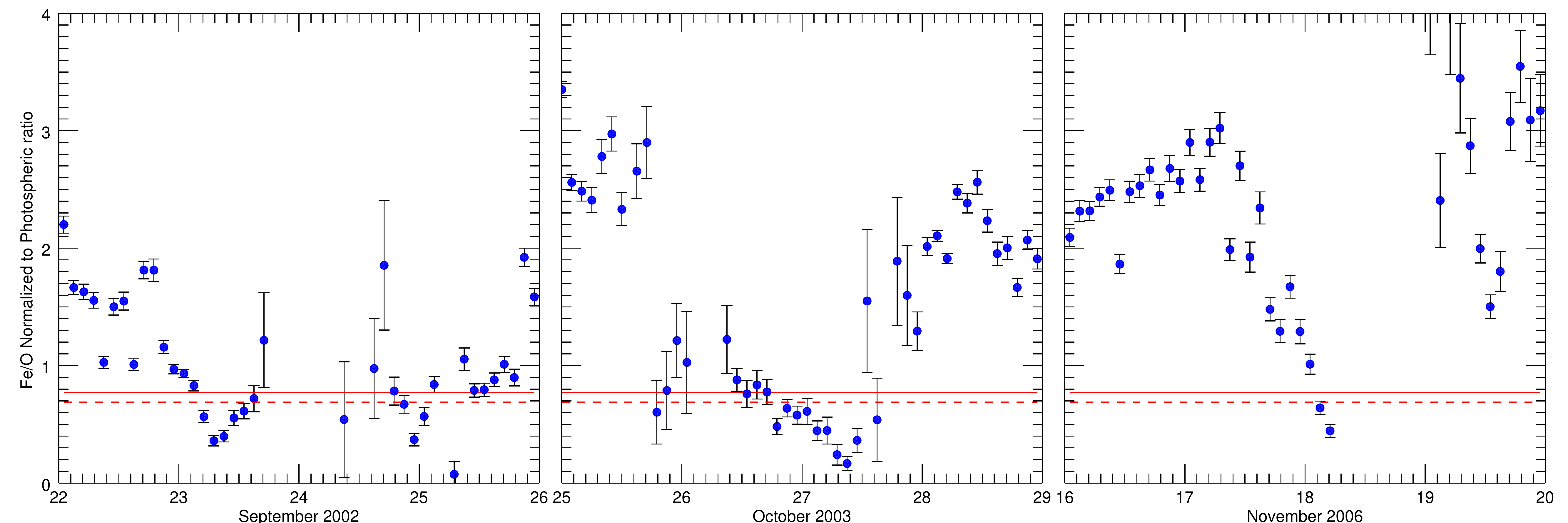}
\caption{ IFIP events detected in the solar wind by ACE/SWICS.
Fe/O abundance ratios measured by SWICS in-situ in the solar wind normalized to their photospheric abundance ratio. Measurement uncertainties are shown by the error bars. The red solid and dashed lines correspond to the same levels as shown in Figure \ref{fig1}. We removed all data flagged as poor/marginal quality or missing due to numerical errors, low statistics, or anomalous operations.
}
\label{fig2}
\end{figure*}

\begin{figure*}
\centering
\includegraphics[width=1.0\textwidth]{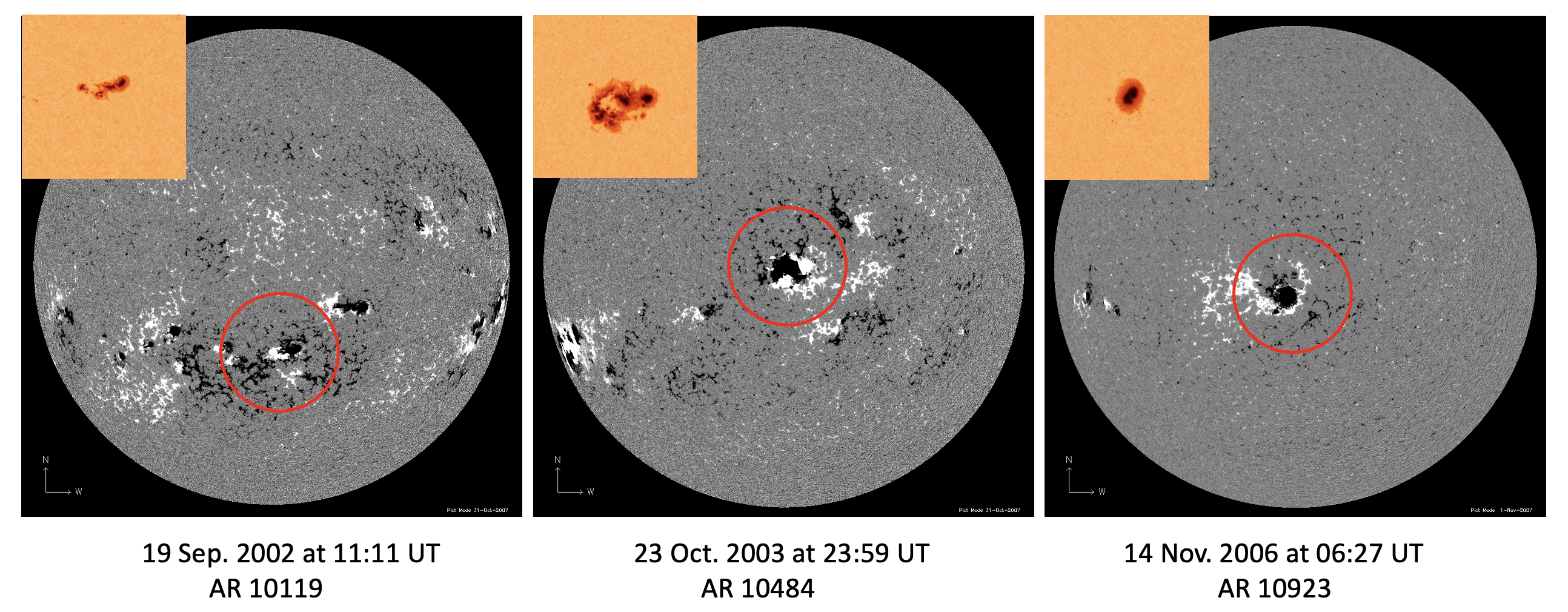}
\includegraphics[width=1.0\textwidth]{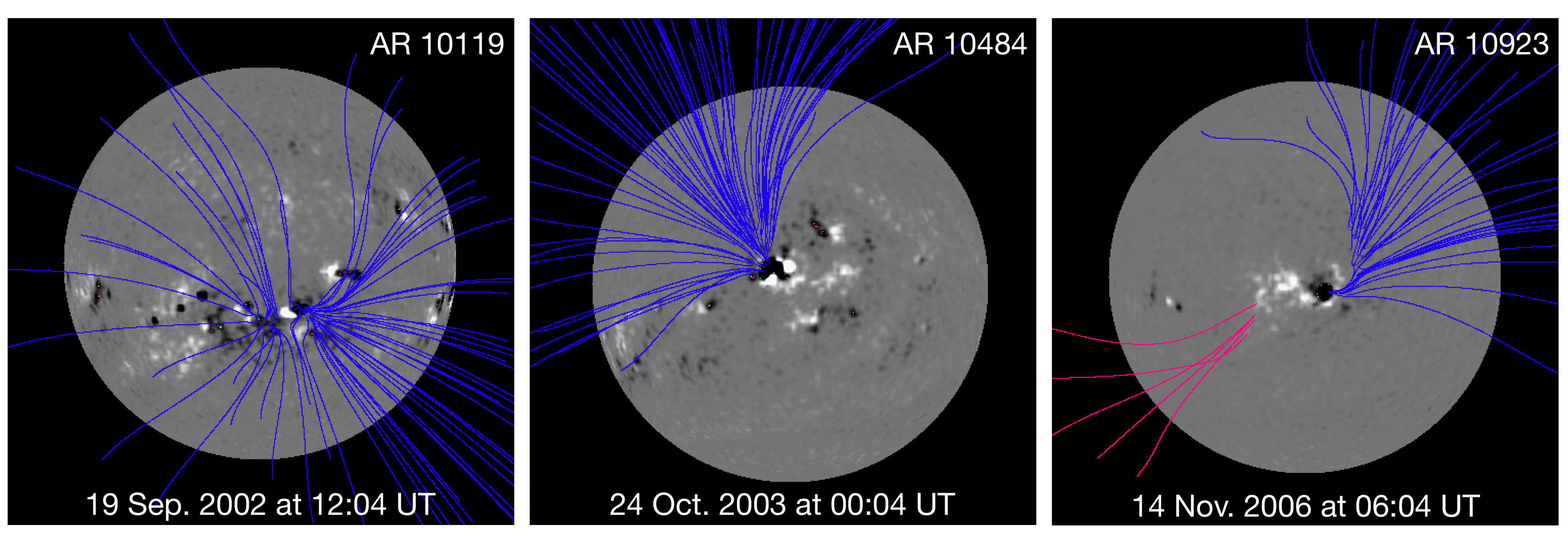}

\caption{ Top panels: SOHO/MDI full solar disk magnetograms with accompanying photospheric continuum images (colored insets).
We show the longitudinal magnetic field in the solar photosphere at the times of the likely departure of the back-mapped solar wind of the events in 2002, 2003, and 2006; with the red circles indicating the likely source regions. These images highlight the complexity of the magnetic field on the Sun during the IFIP events. White indicates positive polarity, and black indicates negative polarity. The images are scaled from -500G to 500G. Bottom panels:
Open coronal magnetic field lines taken from the PFSS model for each active region (AR 10119, 10484, and 10923). The blue field lines indicate negative polarity and the pink field lines represent positive polarity magnetic field. The grayscale images are reduced resolution MDI full disk observations of the longitudinal magnetic field in the solar photosphere with white showing positive polarity and black showing negative polarity, scaled from -200 to 200~G.
}
\label{fig3}
\end{figure*}

The number of IFIP events so far detected remotely on the Sun is small, and they are all associated with solar flares \citep{Doschek2016,Katsuda2020}. Flares are relatively rare in the sense that they are often short duration and occur infrequently. Unlike sampling the constant flow of solar wind, therefore, it would be challenging to observe any signatures of IFIP events in-situ. It is not established, however, that flares are the only source of IFIP composition. We conjectured, therefore, that if the plasma can escape into the heliosphere and form part of the solar wind (an unknown), it might be possible to detect it in-situ if an appropriate instrument fortuitously crosses the appropriate solar wind stream at the right time. It has been pointed out \citep{Peter2018} that a lack of IFIP detection in the solar wind casts some doubt on the reality of the solar IFIP effect, so we felt it worthwhile to pursue. Our best chance would be at a time when potentially more IFIP composition plasma is being produced, and based on the initial solar observations, this would be most likely at times of extreme magnetic complexity, and high activity. 

Here we report observations of three IFIP events in-situ in the solar wind by the Solar Wind Ion Composition Spectrometer \citep[SWICS,][]{Gloeckler1998} on the ACE Observatory \citep{Stone1998}.

\section{Data and Method}
\label{Methods}

The ACE observations we analyze are SWICS 1.1 level 2 version 4.09 and SWICS 2.0 level 3 version 1.1 data \citep{Shearer2014}, and MAG (Magnetometer) level 2 data that we downloaded from the ACE Science Center at Caltech. We felt that examining a low/high-FIP element ratio that shows a large variation in the solar wind would give us the best chance of detecting IFIP events. We therefore focused on the Fe/O ratio since Fe and O have a large separation in FIP. We normalized the SWICS data to the photospheric Fe/O abundance ratio of \cite{Asplund2009} (Fe = 7.50; O = 8.69; Fe/O = 0.0646). The mean error for the absolute Fe/O abundance ratios in the dataset is less than 2\%. These are the statistical uncertainties due to limited count statistics provided by the SWICS team. They do not include any source of systematic error. Such errors can arise from several sources. For example, misidentification of ions, uncertainties in the geometric sensitivity of the instrument etc. \cite{VonSteiger2011} made an assessment of these error sources and argued that the impact on relative abundance measurements, like Fe/O, is not too large (less than 10\% for the largest factor: detector efficiency -- see their Table A1). When searching the database, we identified IFIP events if the normalized abundance ratio was 30\% lower than the photospheric abundance ratio. Compared to the estimates of statistical and systematic uncertainties a 30\% decrease is significant. We chose this relatively large threshold for detection in order to find the most significant events, and also to compensate for the fact that there is a spread in photospheric Fe/O abundance ratios in the recent literature from 0.0575 -- 0.0646 \citep{Asplund2009,Caffau2011,Scott2015}. Under these conditions, the IFIP events are detected regardless of which of these sources is used for the photospheric abundances; larger variations are seen in earlier datasets (see discussion below), but we restrict ourselves to these recent compilations.

We first searched the SWICS daily database between 1998, February 4, and the hardware problem on 2011, August 23. We show a histogram of all the values during this time-period in Figure \ref{fig1}. The histogram peaks at a normalized Fe/O photospheric value of 2.1 and the standard deviation is 0.7 -- assuming a normal distribution. Our IFIP event detection threshold therefore roughly corresponds to a confidence level of $\sim$95\% (i.e. 2$\sigma$). The distribution, however, is in fact right hand skewed to higher values: implying that the standard deviation assuming a normal distribution is overestimated, so the statistical significance of the detections will be underestimated. The initial search found anomalies on five separate occasions in the daily database. As can be seen in Figure \ref{fig1}, these are very rare events. It is worth noting that even photospheric Fe/O ratios are rare in this dataset. Since we define an IFIP detection as 30\% below photospheric Fe/O values, it makes sense to define photospheric Fe/O ratios as 1$\pm$0.3. Analyzing the skewed distribution, we find that $<$10\% of the values are photospheric events. When IFIP events were detected, we then downloaded the more detailed 2-hourly data, but we were only able to confirm the presence of three events in the 2-hourly data. We also examined the SWICS 2-hourly data post-2011, but found no anomalies in that time-period. In all the cases that were detected the data quality is poor at some stage during or after the event, so the exact durations are difficult to determine. In several cases the measurements are flagged as having data missing from the velocity distribution functions (VDF). In some cases the VDFs are completely empty, whereas in other cases the data are flagged as being marginally valid. We excluded all data that were not flagged as clean. In SWICS data larger uncertainties can also occur due to low count rates, which suggests a depletion of the low-FIP element Fe. Whether low-FIP elements are depleted or high-FIP elements are enhanced during the IFIP effect is another topic of interest. The ponderomotive force model due to \cite{Laming2004} suggests the FIP and IFIP effects act solely on low-FIP elements, and there is some evidence in support of this from observations by the EUV Imaging Spectrometer on the Hinode satellite \citep{Brooks2018}. The SWICS measurements we report here are in agreement with that scenario. If the high-FIP element O were enhanced, we would be less likely to encounter difficulties due to low count rates.

There were multiple numbered active regions on disk during the IFIP events; including several extremely large ones. For each IFIP event we also discuss the potential source region and its coronal magnetic field topology using data obtained by the Michelson Doppler Imager \citep[MDI,][]{Scherrer1995} on the Solar and Heliospheric Observatory \citep[SOHO,][]{Domingo1995}. We used full disk observations of the longitudinal (line-of-sight) magnetic field in the solar photosphere. The data were downloaded from the Joint Science Operations Center (JSOC) at Stanford via the Virtual Solar Observatory gateway and have been calibrated with the best available methods. To examine the coronal magnetic field topology we used potential field source surface (PFSS) models \citep{Schatten1969}. For each time-period of interest we generated PFSS extrapolations using the package  available in SolarSoftware \citep{Schrijver2003}. This package accesses a database of potential field models constructed from the MDI photospheric magnetic field data. Field lines are traced out from the solar surface with those reaching the source surface designated as open.

\section{Results and discussion}

The three events we describe occurred in September 2002, October 2003, and November 2006. 
We show the Fe/O abundance ratios, normalized to their photospheric ratios, in Figure \ref{fig2}, and  
full disk images of the longitudinal magnetic field at the back-mapped time of these detections in Figure \ref{fig3} along with potential field extrapolations indicating the locations of open magnetic field at the times of the likely departure of the ACE measured speed back-mapped solar wind. The figure highlights the three potential active region sources of the events we detected (AR 10119, 10484, and 10923), and we highlight the extrapolated open field lines connected to these regions.

The clearest and most significant event we found occurred on October 26--27, 2003. This was during the famous Halloween storms, when numerous major flares ($>$ M5 class) occurred between 19 October and 20 November, including four of the ten strongest X-class flares recorded this century, and the most powerful (X45) flare of the space age; although the source region of the most intense flares was AR 10486 not AR 10484, which, however, also produced a few X-class and several M-class flares. The Fe/O ratio also reached record highs during this period, so solar conditions were at their most extreme in the days surrounding the time of this event. The Fe/O ratio dropped below photospheric levels from $\sim$11\,UT on the 26th and recovered to nominal values some time after 11\,UT on the 27th. The event was therefore sustained for around 24 hours.

\begin{figure}[h]
\centering
\includegraphics[width=0.5\textwidth]{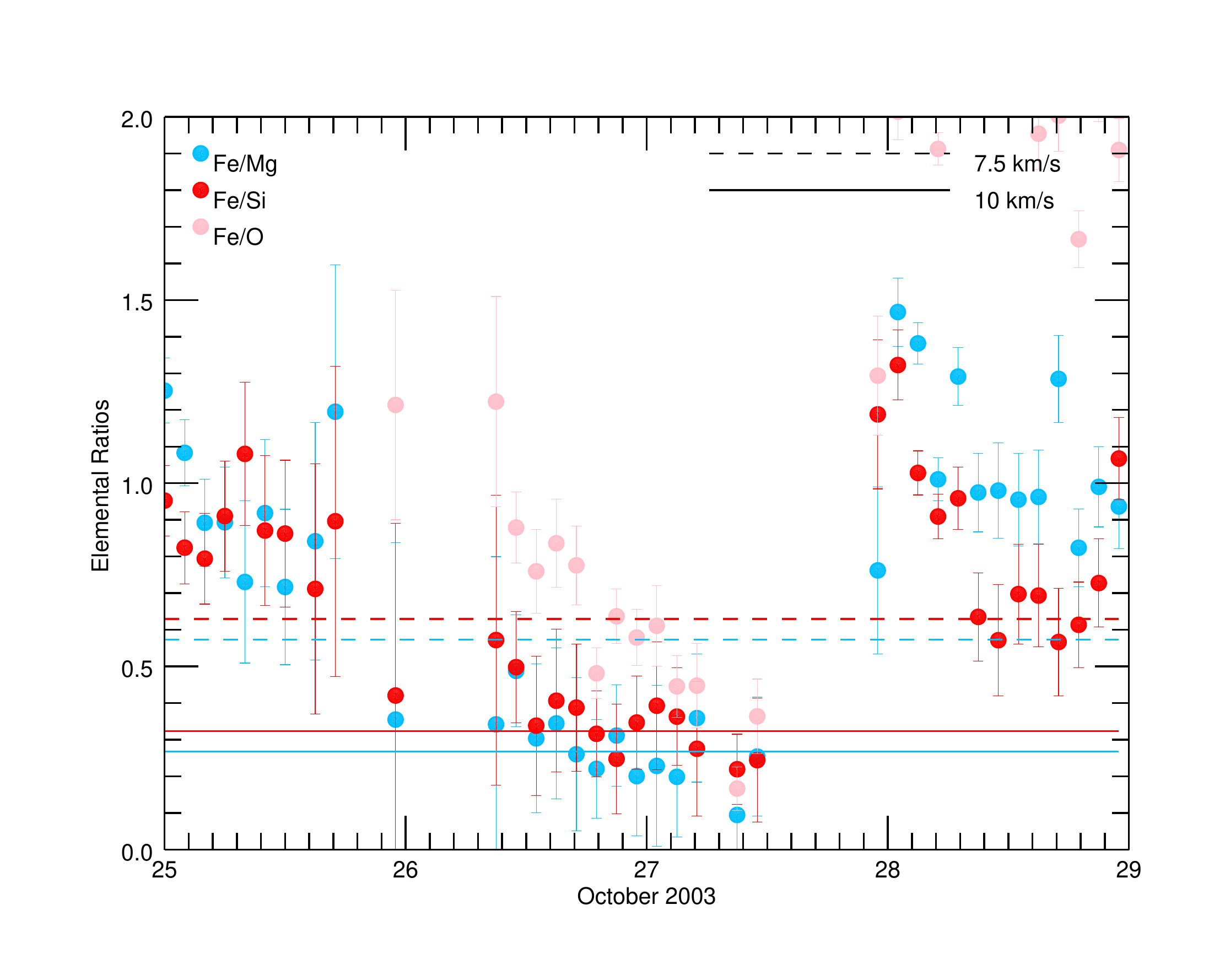}
\caption{Elemental abundance ratios compared to the ponderomotive force model.
Fe/O, Fe/Si, and Fe/Mg abundance ratios measured by SWICS in-situ in the solar wind during the October 2003 IFIP event and normalized to their photospheric abundance ratios. 
We compare the measurements to theoretical values predicted from the ponderomotive force model of IFIP due to \cite{Laming2015}. 
We show the theoretical predictions for two different values of the fast mode wave amplitude in the models. 
Measurement uncertainties are shown by the error bars.
}
\label{fig4}
\end{figure}

The September 2002 event was also potentially of long duration but was less clear. The Fe/O ratio dropped below photospheric levels around 23\,UT on the 22nd, but only reached values significant enough for our detection (see section \ref{Methods}) around 5\,UT on the 23rd. The data are highly uncertain or missing between 17\,UT on the 23rd and 19\,UT on the 24th before IFIP composition is detected again until $\sim$1\,UT on the 25th and photospheric composition thereafter. It is unclear whether the two detections are part of the same event, but they were also associated with high solar activity. Two 
M-class flares occurred a few days before, and there were an unusually large number of active regions ($\geq15$) on disk at the solar wind velocity back-mapped time (see Figure \ref{fig3}).

The November 2006 event was only detected for a short duration. A normal FIP effect was observed in the solar wind one day prior to the event, but the Fe/O ratio trended downward over the preceding 24 hours and hit sub-photospheric levels for 2-6 hours around 4\,UT on the 18th. The data are highly uncertain after that time, but if this is due to low count rates for Fe then it is plausible that this event was also long duration. Flaring activity was not significant during this event, but, at the back-mapped time there was an active region with a very large leading spot in the middle of the solar disk (see Figure \ref{fig3}). The average magnitude of the IFIP effect is about a factor of 2 in these three examples. 

There may be other ways that similar events could be detected in the solar wind. We searched the literature for older observations and found some interesting historical evidence in support of these detections from SOHO/CELIAS. \cite{Aellig1999} studied the Fe/O abundance ratio over an eighty day period near solar minimum in 1996, $\sim$1 June 1996--19 Aug 1996, and show distributions of values for different wind speeds extending down to photospheric abundances ($\sim$1.4 in their Figure 3; which shows log Fe/O ratios). At the time of their calculations, however, the accepted photospheric Fe/O abundance ratio dated back to 1989 \citep{Anders1989}, and is a factor of 1.8 lower than recent estimates from the same group \citep{Asplund2009}. This means that the tail of the distribution of Fe/O ratios for the slow wind ($<$360\,km s$^{-1}$) in the CELIAS data, for example, is showing an IFIP effect of a factor of 2, albeit that this part of the distribution also represents rare cases.

One potentially very important consequence of the detection of these events in the solar wind is that they allow new examination of the properties of MHD waves, that can heat and accelerate the solar wind, using other in-situ measurements. In Figure \ref{fig4} we make an initial comparison of the SWICS observations from October 2003 with the ponderomotive force model of the IFIP effect \citep{Laming2015}. In this model, chromospheric acoustic waves propagating from below mode convert to become fast-mode waves that then reflect downward. This leads to a downward directed ponderomotive force that depletes the region of low-FIP elements. We show the SWICS Fe/O, Fe/Si, and Fe/Mg abundance ratios during this event, plotted against the predicted ratios from the IFIP effect model. The predictions are made based on different values of the fast mode wave amplitude at the layer where magnetic and plasma pressure is approximately equal. The observed results are broadly consistent with fast mode wave amplitudes of 7.5--10\,km s$^{-1}$ for all three events. These values are similar to the lower end of wave amplitudes measured previously in the chromosphere and corona, but are still strong enough to accelerate the solar wind if the energy flux transmission coefficient is high enough \citep{DePontieu2007}.

\begin{figure}[h]
\centering
\includegraphics[viewport=0 130 512 1536,clip,width=0.45\textwidth]{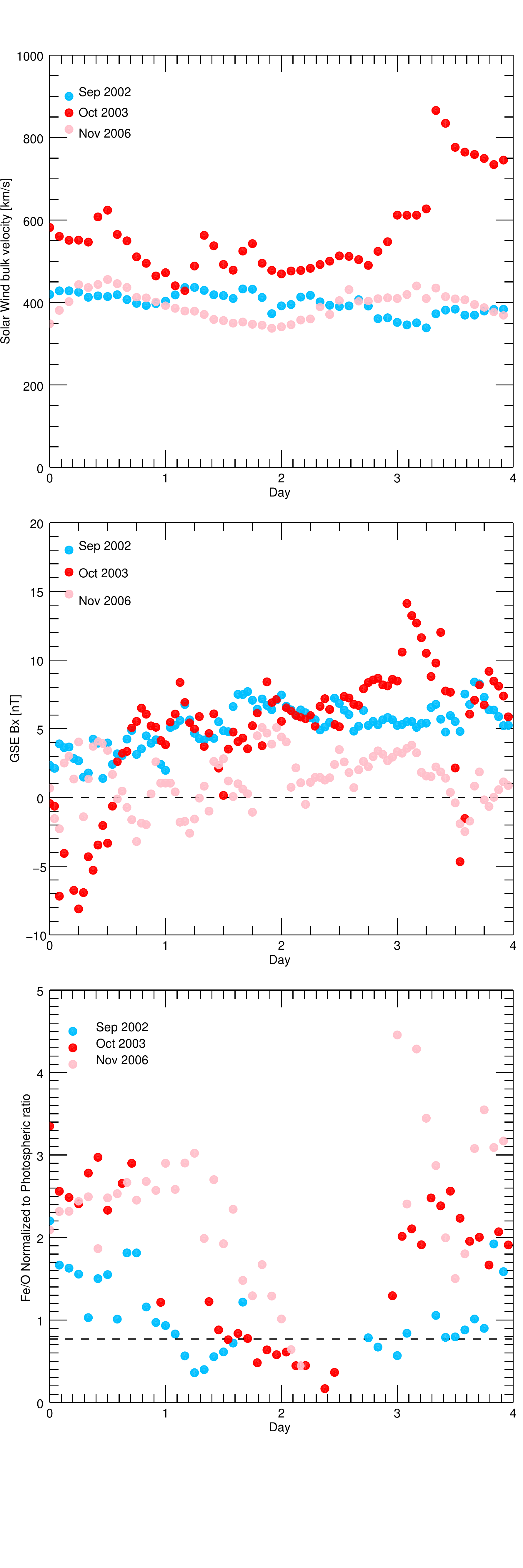}
\caption{  Solar wind velocity, interplanetary magnetic field direction, and Fe/O ratios during the IFIP events.
Top panel: He++ velocity measured by SWICS in-situ around the time period of the three IFIP events shown in Figure \ref{fig2}. 
Middle panel: 
B$_x$ measured by ACE/MAG in-situ around the time period of the three IFIP events shown in Figure \ref{fig2}. 
Bottom panel: The three IFIP events of Figure \ref{fig2} plotted for ease of cross-reference.
We show all the data on a common time-scale that corresponds to the ranges of Figure \ref{fig2}
}
\label{fig5}
\end{figure}

Using another in-situ measurement we also show that all three IFIP events are detected in the slow solar wind. Figure \ref{fig5} (top panel) shows the bulk solar wind speed in the time-periods covering these detections. In all cases the wind speed is below $\sim$600\,km s$^{-1}$ in the days preceding the events, while the plasma travelled to the L1 Lagrange point. The origin of the slow solar wind is still under debate, and active regions have been suggested as one possible source \citep{Liewer2004}. There is a difficulty, however, in using plasma composition measurements to definitively identify the sources. This is because an enhanced slow wind composition may be observed in multiple active regions on the solar disk at the same time, and even in different features such as coronal loops and outflows within the same active region \citep{Brooks2011}. The discovery of IFIP composition plasma in the slow wind implies that specific magnetically complex areas that produce this within active regions can also be slow solar wind sources. To be clear, solar observations have thus far only detected IFIP composition plasma above sunspots in complex flaring regions. Our observations indicate that complex sunspots are a previously unrecognized source of the slow solar wind. 

Our ACE-measured solar wind speed back-mapped times, however, all single out one respective active region as the most likely source of the IFIP-composition slow solar wind. All three active regions have open field lines, the magnetic polarities of which match the polarities of the slow wind where the IFIP composition was detected (as we show in Figure \ref{fig5} - middle panel). The figure shows the interplanetary magnetic field direction, B$_x$ in Geocentric Solar Ecliptic (GSE) coordinates, measured by ACE/MAG. The data are plotted on a common time-scale as in the upper panel. For the September 2002 event (sky blue dots), B$_x$ is always positive i.e. directed sunward; which matches the negative polarity 
(opposite convention) of the open magnetic field observed in the active region. For the October 2003 event, B$_x$ is initially negative, but has turned positive by the time the IFIP effect is detected. The November 2006 measurements are more variable, with B$_x$ spreading positive and negative values as the FIP bias trends downward (November 17; day 1 in Figure \ref{fig5}), but is again predominantly positive when the IFIP effect is detected (November 18; day 2 in Figure \ref{fig5}).

Only one of the three active regions (AR 10484), however, is a highly complex emerging AR similar to those in which IFIP composition was detected previously \citep{Baker2020}. The second potential source region is a medium-sized AR (AR 10119), which was in its emergence phase when it was linked to IFIP-composition solar wind. The third AR (AR 10923) has its large decaying leading spot to make it stand out. These two latter active regions broaden the class of solar active regions which may produce IFIP coronal plasma composition to cases that are less unusual, indicating that localised IFIP production in active regions may be more common than previously thought. Active regions with high magnetic flux density in their emergence phase may be good candidates for IFIP plasma production. Observing clear signals of this in the slow solar wind would be an unlikely event, however, since a number of conditions would have to be met. For example, there must be open magnetic field in the vicinity of the sunspot umbra where the usually highly localized IFIP plasma is observed - or an alternative escape pathway. Furthermore, the IFIP composition plasma may be localised within the AR embedded in a coronal composition dominated environment, so plasma mixing with surrounding coronal composition material as the solar wind forms may dilute the signature and mask most of these events before they reach the observing spacecraft. Since these events are also of relatively short duration (compared to the active region lifetime), there is a reduced chance of it being at the footpoint of the slow-wind stream that connects to the spacecraft; which is itself a low probability event.

The rarity of the IFIP events of course implies that their source is only a very minor contributor to the slow solar wind. They are potentially very valuable, however, for remote-sensing and in-situ connection modelling, precisely because they are rare and localized to specific regions. The detections by SWICS suggest that future missions such as Solar Orbiter may also observe IFIP events, and it may be easier to link these to specific footpoints of the wind connecting to the spacecraft because there may be only one location producing IFIP composition plasma. Very detailed observations of the source region will then become possible, allowing refinement of the model predictions that are currently used.

The discovery of these events also confirms that IFIP composition plasma can escape into the heliosphere. Although there is no theoretical reason to think otherwise, it is important to stress that this was unknown before the detection, and has implications for stellar observations. The fact that there is no impediment to this plasma flowing into the solar wind implies that solar-like stars with IFIP composition coronae, such as M-dwarfs, have stellar winds that are also dominated with IFIP composition plasma; very different from the Sun. Furthermore, any gradual solar energetic particle (SEP)-like events on these stars will also have an IFIP composition, since the seed population comes from the corona, with implications for exoplanet atmospheres.

All the previous spectroscopic results are necessarily dependent on the quality of the underlying atomic physics. In the case of the Sun, only a limited number of specific Ar and Ca spectral lines have been used to detect the IFIP effect, and the particular Ca line (Ca XIV 193.874\,\AA) has shown unusual width behavior in solar observations \citep{Brooks2016}. Modeling assumptions, such as ionization equilibrium, also introduce uncertainties. For the M-dwarfs, there are generally no measurements of their photospheric abundances \citep{Laming2015}, so the IFIP effect is computed relative to solar photospheric abundances. This means that if their photospheric elemental abundance distributions are different than that of the Sun, as their coronal and apparently now wind compositions appear to be, then the IFIP effect could be smaller or even non-existent. Our detection of the inverse FIP effect by particle counting techniques in the solar wind lends support to all the inverse FIP effect findings by completely different spectroscopic methods.

\acknowledgments 
We thank the anonymous reviewer for several detailed comments and suggestions that improved the manuscript.
The work of D.H.B. and H.P.W. was funded by the NASA Hinode program. D.B. is funded under STFC consolidated grant number
ST/S000240/1 and L.v.D.G. is partially funded under the same grant. L.v.D.G. acknowledges the Hungarian National Research, Development and Innovation Office grant OTKA K-131508. The work of S.L.Y. was carried out with support from the UKRI SWIMMR Aviation Risk Modelling (SWARM) grant.
We thank the ACE SWICS and MAG instrument teams and the ACE Science Center for providing the ACE data.

\end{document}